\documentclass{article}
\usepackage{spconf,amsmath,graphicx}
\usepackage{multirow}
\usepackage{amssymb}
\usepackage{pifont}
\usepackage{xcolor}
\usepackage{url}
\usepackage{svg}

\usepackage{enumitem}
\setlist{nosep, leftmargin=14pt}

\usepackage{mwe} 

\title{Is autoencoder truly applicable for 3D CT Super-Resolution?}
\name{Weixun Luo$^{\star \dagger}$ \qquad Xiaodan Xing$^{\dagger}$ \qquad Guang Yang$^{\dagger \ddagger}$}
\address{$^{\star}$ Department of Bioengineering, Imperial College London, London, UK \\
         $^{\dagger}$ National Heart and Lung Institute, Imperial College London, London, UK\\
         $^{\ddagger}$ Cardiovascular Research Centre, Royal Brompton Hospital, London, UK}
\begin{document}
%
\maketitle
\begin{abstract}
Featured by a bottleneck structure, autoencoder (AE) and its variants have been largely applied in various medical image analysis tasks, such as segmentation, reconstruction and de-noising. Despite of their promising performances in aforementioned tasks, in this paper, we claim that AE models are not applicable to single image super-resolution (SISR) for 3D CT data. Our hypothesis is that the bottleneck architecture that resizes feature maps in AE models degrades the details of input images, thus can sabotage the performance of super-resolution. Although U-Net proposed skip connections that merge information from different levels, we claim that the degrading impact of feature resizing operations could hardly be removed by skip connections. By conducting large-scale ablation experiments and comparing the performance between models with and without the bottleneck design on a public CT lung dataset , we have discovered that AE models, including U-Net, have failed to achieve a compatible SISR result ($p<0.05$ by Student's \textit{t}-test) compared to the baseline model. Our work is the first comparative study investigating the suitability of AE architecture for 3D CT SISR tasks and brings a rationale for researchers to re-think the choice of model architectures especially for 3D CT SISR tasks. The full implementation and trained models can be found at: https://github.com/Roldbach/Autoencoder-3D-CT-SISR
\end{abstract}
\begin{keywords}
Autoencoder, super-resolution, CT
\end{keywords}
%

\begin{figure*}[ht]
    \centering
    \resizebox{2.0\columnwidth}{!}{%
        \includegraphics{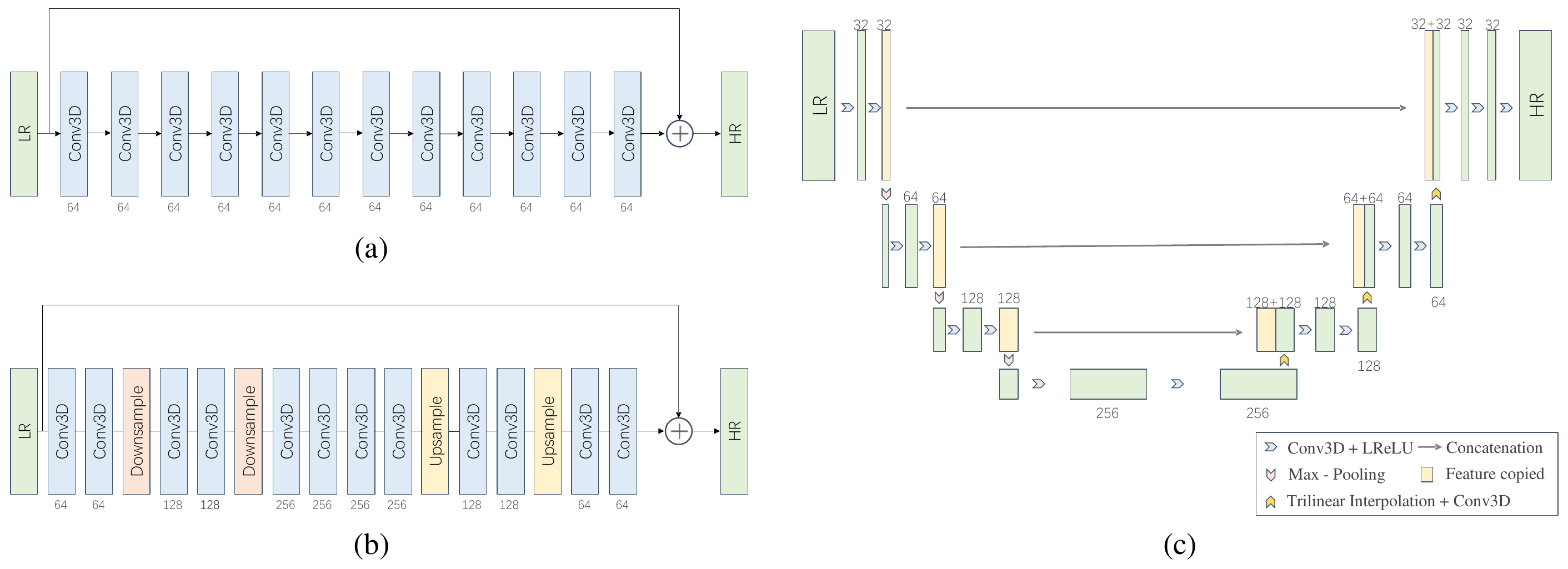}
    }
    \caption{Architectures of different models used in this study. (a) Plain CNN; (b) AE models; (c) U-Net.}
    \label{fig:architecture}
\end{figure*}

\section{Introduction}
\label{sec:introduction}
High resolution (HR) volumetric data generated by Computed Tomography (CT) can capture small structures and provide detailed textural information about human anatomy and pathology, thus facilitate the diagnostic procedure . However, the acquisition of HRCT data requires exposure to high-dose radiation, which can bring potential health risks to patients. More importantly, HRCT data are often downsampled by increasing the slice interval to reduce the intrinsically high storage requirement. Unfortunately, the downsampled data are less likely to be re-used in subsequent image analysis that requires high-quality input.

To address these dilemmas, single image super-resolution (SISR) has attracted increasing attention as it requires only one low resolution (LR) instance to reconstruct the HR counterpart without affecting the raw data acquisition. Compared with 2D SISR, 3D SISR is considerably more challenging. First, the size of 3D volumetric data easily leads to memory bottlenecks and prolonged training time. Moreover, 3D data contain vastly more contextual and structural details that impose additional difficulties for 3D SISR model training. Finally, the use of 3D convolutional layers inevitably necessitates a much higher number of parameters than in 2D, so the size of the model must be very carefully considered.

Among 3D super resolution models, a popular memory efficient solution is the utilization of the autoencoder (AE) architecture \cite{C15}, where feature maps could be substantially downsampled to a great extent in the middle of the model. By downsampling feature maps, the number of parameters and the time required for optimization are largely reduced. Typically, common downsampling layers include pooling \cite{C16}, stride convolution \cite{C17} and interpolation \cite{C18}. Besides of its applications on SR tasks, the AE structure has also been successfully applied in other tasks such as segmentation \cite{C19} and detection \cite{C20}.

In this paper, however, we show that \textbf{models utilizing AE, including U-Net, have substantial limitations for 3D CT SISR.} Specifically, we compare various AE models with the baseline model, and provide statistically significant evidence to show that AE structures cause an unrecoverable loss of information during the data processing and potentially increases the training difficulty. We also demonstrated that skip connections and the feature map concatenations in U-Nets may mitigate the negative effect caused by the feature map down sampling, but they can not fully compensate for the information loss. 

\section{Methods}
\label{sec:methods}
In this section, we describe the implementation of each of the models used in the experiments. We first build the baseline model, “Plain CNN”, upon the simplest backbone to avoid any possible benefits brought by the architecture itself. All AE models are then constructed by introducing resizing layers into Plain CNN to ensure the feature resizing operation is the only control variable in the comparison experiments. Moreover, we adapt the U-Net, primarily to investigate the effect of skip connections on the performance of AE models.

\subsection{Plain CNN}
\label{ssec:plainCNN}
Fig. \ref{fig:architecture}(a) illustrates the structure of Plain CNN, which consists of 12 basic building blocks connected in series. Each block consists of a standard 3D convolutional layer with 64 filters of size $3\times3\times3$ and a Leaky Rectified Linear Unit (Leaky ReLU) with slope of 0.1 as the nonlinear activation function. To avoid resizing effects, we keep the same dimension for all feature maps by setting the stride to 1 for every convolutional layer and adding zero padding before convolution. We abandon the use of Batch Normalisation (BN) layer within the model. According to \cite{C10}, BN not only occupies too much memory but also discards valuable feature range flexibility in SR. Finally, we apply the global residual learning as suggested by \cite{C23} to ease training and prevent the gradient vanishing problem.

\subsection{Autoencoder}
\label{ssec:autoencoder}
Given the baseline model, we insert a downsampling layer after every 2 building blocks and its corresponding upsampling layer symmetrically, turning the model into the AE architecture shown in Fig. \ref{fig:architecture}(b). Each of these layers resizes the feature map in all dimensions by a factor of 2 and the channel number is adjusted to compensate for this effect. We consider 2 options for the downsampling layer: 1) max-pooling with a filter of size $2\times2\times2$, stride of 2 and dilation of 1; and 2) 3D convolutional layer with a filter of size $3\times3\times3$, setting stride to 2 and padding to 1. To prevent any checkerboard artifacts generated by the transpose convolution \cite{C24} during the upsampling operation, we use trilinear interpolation to resize feature maps followed by the standard convolution.

\subsection{U-Net}
\label{ssec:unet}
Our implementation of the U-Net model is based on \cite{C25}, which is shown in Fig. \ref{fig:architecture}(c). We replace all 2D convolutional layers and pooling layers with their corresponding 3D versions without changing their configurations. For the reasons mentioned above, BN layers are not used and all transpose convolution operations are substituted with “trilinear interpolation + convolution” in the decoder pathway. To ensure a relatively fair comparison with AE models, we simplify the U-Net to have a similar model size and depth by: 1) setting the channel number in the convolutional layer at the first level to 32 instead of 64; 2) reducing the level of hierarchical feature maps from 5 to 4.

\begin{table}[t]
    \centering
    \resizebox{1.0\columnwidth}{!}{%
    \begin{tabular}{ccccc} 
    \hline
    Scale               & Methods    & PSNR                                   & SSIM                                      & RMSE                                   \\ 
    \hline
    \multirow{4}{*}{$\times2$} & AE-Maxpool & 35.14 (2.50)*                          & 0.9649 (0.0062)*                          & 4.64 (1.20)*                           \\
                               & AE-Conv    & 35.38 (2.53)*                          & 0.9641 (0.0072)*                          & 4.52 (1.21)*                           \\
                               & U-Net      & 39.40 (1.23)*                          & 0.9795 (0.0040)*                          & 2.76 (0.40)*                           \\
                               & Plain CNN  & \textbf{43.52 (0.95)}{\color{white} *} & \textbf{0.9839 (0.0049)}{\color{white} *} & \textbf{1.71 (0.19)}{\color{white} *}  \\ 
    \hline
    \multirow{4}{*}{$\times4$} & AE-Maxpool & 26.89 (2.51)*                          & 0.8711 (0.0213)*                          & 11.99 (3.12)*                          \\
                               & AE-Conv    & 25.49 (1.29)*                          & 0.8741 (0.0207)*                          & 13.71 (2.1)*                           \\
                               & U-Net      & 29.31 (1.73)*                          & 0.9131 (0.0112)*                          & 8.92 (1.94)*                           \\
                               & Plain CNN  & \textbf{34.51 (0.65)}{\color{white} *} & \textbf{0.9345 (0.0130}){\color{white} *} & \textbf{4.81 (0.35)}{\color{white} *}  \\ 
    \hline
    \multirow{4}{*}{$\times8$} & AE-Maxpool & 24.20 (1.86)*                          & 0.7890 (0.0320)*                          & 16.08 (3.22)*                          \\
                               & AE-Conv    & 23.34 (1.59)*                          & 0.7838 (0.0306)*                          & 17.63 (2.88)*                          \\
                               & U-Net      & 30.24 (1.20)*                          & \textbf{0.8663 (0.0220){\color{white} *}} & 7.92 (1.19)*                           \\
                               & Plain CNN  & \textbf{31.03 (1.25)}{\color{white} *} & 0.8644 (0.0226){\color{white} *}          & \textbf{7.23 (0.96)}{\color{white} *}  \\ 
    \hline
                               &            &                                        &                                           &                                        \\ 
    \hline
    \multirow{4}{*}{$\times2$} & AE-Maxpool & 33.66 (2.16)*                          & 0.9539 (0.0094)*                          & 5.44 (1.14)*                           \\
                               & AE-Conv    & 32.25 (3.23)*                          & 0.9468 (0.0108)*                          & 6.63 (2.13)*                           \\
                               & U-Net      & 37.52 (1.43)*                          & 0.9770 (0.0035)*                          & 3.44 (0.63)*                           \\
                               & Plain CNN  & \textbf{43.16 (1.02)}{\color{white} *} & \textbf{0.9837 (0.0048)}{\color{white} *} & \textbf{1.79 (0.21)}{\color{white} *}  \\ 
    \hline
    \multirow{4}{*}{$\times4$} & AE-Maxpool & 23.52 (1.06)*                          & 0.8434 (0.0260)*                          & 17.12 (1.96)*                          \\
                               & AE-Conv    & 25.70 (2.54)*                          & 0.8420 (0.0290)*                          & 13.77 (3.57)*                          \\
                               & U-Net      & 33.45 (0.97)*                          & 0.9287 (0.0152)*                          & 5.45 (0.61)*                           \\
                               & Plain CNN  & \textbf{36.54 (0.69)}{\color{white} *} & \textbf{0.9422 (0.0134)}{\color{white} *} & \textbf{3.81 (0.30)}{\color{white} *}  \\ 
    \hline
    \multirow{4}{*}{$\times8$} & AE-Maxpool & 22.36 (1.48)*                          & 0.7562 (0.0302)*                          & 19.70 (2.98)*                          \\
                               & AE-Conv    & 20.86 (1.42)*                          & 0.7400 (0.0368)*                          & 23.38 (3.55)*                          \\
                               & U-net      & 27.21 (2.14)*                          & 0.8538 (0.0198)*                          & 11.50 (3.36)*                          \\
                               & Plain CNN  & \textbf{31.35 (1.34)}{\color{white} *} & \textbf{0.8757 (0.0197)}{\color{white} *} & \textbf{6.98 (0.99)}{\color{white} *}  \\
    \hline
    \end{tabular}
    }
    \caption{Quantitative comparisons of different models. Best results are shown in \textbf{Bold}. * indicates statistically significant evidence to support the difference with Plain CNN. Top: Mean (STD) using trilinear interpolation in LR generation; Bottom: Mean (STD) using same insertion in LR generation.}
    \label{tab:table1}
\end{table}

\begin{table}[t]
    \centering
    \begin{tabular}{ccc} 
    \hline
    Methods    & \#Parameter (M) & Inference Time (s)  \\ 
    \hline
    AE-Maxpool & 6.88            & 18.46               \\
    AE-Conv    & 7.44            & 18.62               \\
    U-Net      & 5.30            & 10.04               \\
    Plain CNN  & 1.11            & 33.87               \\
    \hline
    \end{tabular}
    \caption{Computational comparisons of different models.}
    \label{tab:table2}
\end{table}

\section{Experiments}
\label{sec:experiment}

\subsection{Dataset}
\label{ssec:dataset}
We use the AAPM-Mayo Clinic Low-Dose CT Grand Challenge Dataset \footnote{\url{https://www.aapm.org/grandchallenge/lowdosect/}} provided by Mayo Clinic for model training and testing. From the CT scans collected from 140 patients, we use 48 chest CT scans and further split these into 37 training, 5 validation and 6 test volumes. The data for each patient consists of normal-dose CT (NDCT) scans and the corresponding synthetic low-dose CT (LDCT) scans with additional Poisson noise. Only NDCT are used as ground truth HR data in the experiments. All volumes within the dataset contain an uneven number of slices of size $512\times512$ with 1.5mm thickness. To enable a feasible training time and solve the memory limitation, we pre-downsampled each slice from $512\times512$ to $256\times256$ using bilinear interpolation.

\subsection{LR Data Generation}
\label{ssec:LRDataGeneration}
Our LR data are degraded from ground truth HR data by the following steps: 1) Truncate the leading and trailing slices evenly so that the dimension of each volume can perfectly fit the non-overlapping patch extraction algorithm; 2) Downsample the volume in the axial direction by removing slices at a constant interval; 3) Clip all HU values into the range [-1024, 1476] and normalize them to [0,1];  and 4) Upsample the volume to its original dimension either by trilinear interpolation or by inserting the same slice at the previous position.


\subsection{Evaluation Metrics}
\label{ssec:evaluationMetrics}
The performance of each model is evaluated by using: Peak Signal-To-Noise-Ratio (PSNR); Structural Similarity Index (SSIM); and the Root Mean Square Error (RMSE). Both PSNR and RMSE focus on the pixel-level error between the reconstructed volume and the ground truth label while SSIM is more suitable for reflecting the structural correspondences.

\begin{figure*}[t]
    \centering
    \resizebox{2.0\columnwidth}{!}{%
        \includegraphics{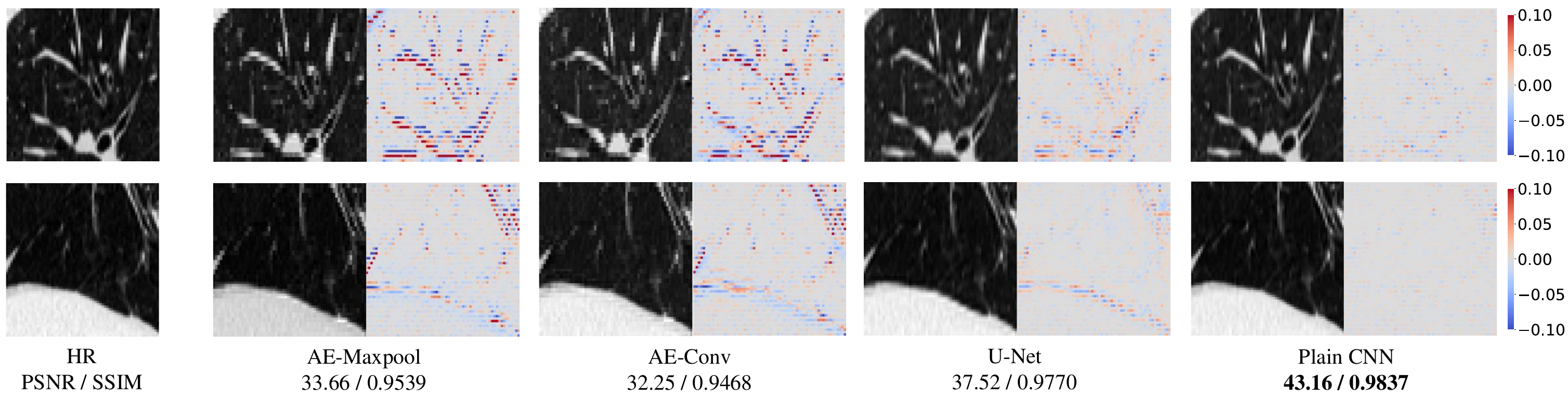}
    }
    \caption{Visual comparisons of different models using same insertion in LR generation under $\times2$ scaling factor.}
    \label{fig:result}
\end{figure*}

\subsection{Statistical Test}
\label{ssec:statisticalTest}
We use Shapiro-Wilk test to check the normality of the difference in the image quality between each AE model and Plain CNN and paired Student's \textit{t}-test to determine whether there is statistically significant evidence to support this difference. We use Wilcoxon signed-rank test instead when the normality of the sample cannot be satisfied. We set the significance level to 0.05 in all statistical tests.

\section{Results and Discussion}
\label{sec:resultsAndDiscussion}
In Table \ref{tab:table1}, we compare the quantitative performance of all AE models with Plain CNN under $\times2$, $\times4$, and $\times8$ scaling factors. Next, we present the total number of parameters and the average inference time used by one volume for every model in Table \ref{tab:table2}. Finally, We show visual comparisons of different models in Fig. \ref{fig:result}.

The results in Table \ref{tab:table1} show that AE is unsuitable for 3D CT SISR. It can be seen that there is an obvious performance drop for all AE models compared with Plain CNN in almost all cases. At the same time, there also exists statistically significant evidence to support this performance drop in almost every comparison experiment. We contend that the main reason for the performance drop is the resizing operation within AE since this is the only architectural difference between AE models and the baseline model. Surprisingly, those results also reveals the fact that skip connections, which are designed to benefit U-Net by increasing the high-resolution feature re-usability, cannot fully compensate for the aforementioned performance gap. From Table \ref{tab:table2}, AE models have a lower computational cost than Plain CNN reflected by the reduced average inference time, but at the cost of a larger model size.

We show comparisons between AE models and Plain CNN visually in Fig. \ref{fig:result}. These results again argue that AE, including U-Net, is not suitable basis for 3D SISR tasks. It is clearly shown that AE models generate noticeable artifacts around edges and significant deviations in regions with abundant textures. This shows the loss of diagnostically important information due to resizing. In contrast, Plain CNN can recover comparatively more natural textures and smoothing edges and produce results that are almost indistinguishable from the ground truth HR data.

\section{Conclusion}
\label{sec:conclusion}
In this paper, we have shown that AE models, including U-Net, are unsuitable for 3D CT SISR primarily due to the information loss in feature resizing operations. We have presented a carefully designed set of experiments, adjusting model architectures for a fair comparison. We have evaluated the models on a publicly available CT lung dataset and have concluded that although AE models can achieve faster inference, they do so at the cost of inferior performance compared to the baseline CNN.

In future work we hope to explore whether the use of other loss functions such as Mean Absolute Error (MAE) and perceptual loss could compensate for the aforementioned information loss. We also plan to investigate the performance of AE-based generative adversarial networks (GAN).

\section{Compliance with ethical standards}
This research study was conducted retrospectively using human subject data made available in open access by 2016 Low-dose CT AAPM Grand Challenge. Ethical approval was not required as confirmed by the license attached with the open access data.
\label{sec:compliance}

\section{acknowledgments}
\label{sec:acknowledgments}
This study was supported in part by the ERC IMI (101005122), the H2020 (952172), the MRC (MC/PC/21013), the Royal Society (IEC/NSFC/211235), the Imperial College Undergraduate Research Opportunities Programme (UROP), the NVIDIA Academic Hardware Grant Program, the SABER project supported by Boehringer Ingelheim Ltd, NIHR Imperial Biomedical Research Centre (RDA01), and the UKRI Future Leaders Fellowship (MR/V023799/1).

\bibliographystyle{IEEEbib}
\bibliography{refs}

\end{document}